\documentclass[prc,twocolumn,showpacs,superscriptaddress,preprintnumbers,amsmath,amssymb,nofootinbib]{revtex4-1}
\usepackage{bm}
\usepackage{dcolumn}
\usepackage{multirow}
\usepackage{ifpdf}
\usepackage{url}
\usepackage[utf8]{inputenc}

\ifpdf
  \usepackage{graphicx}     %   usepackage without driver option
  \usepackage{hyperref}
\else     % For (p)LaTeX + dvipdfmx
  \usepackage[dvipdfmx]{graphicx}     %   usepackage with driver option
  \usepackage[dvipdfmx]{hyperref}
\fi

\newcommand\pythia{\textsc{Pythia}}

\newcommand{\BV[1]}{\mbox{\boldmath ${#1}$}}

\usepackage[normalem]{ulem}  % \sout{old text} for strikeout
\usepackage[dvips]{color} % For blue in-text comments and additions

\renewcommand\sout{\bgroup \color{red} \ULdepth=-.5ex \ULset}

\hypersetup{
colorlinks=true,
linkcolor=blue,
citecolor=blue,
urlcolor=blue
}

\begin{document}

\title{
New approach to initializing hydrodynamic fields\\
 and mini-jet propagation
in quark-gluon fluids
}

\author{Michito Okai}
\email{michito0605@eagle.sophia.ac.jp}
\affiliation{%
Department of Physics, Sophia University, Tokyo 102-8554, Japan
}
\author{Koji Kawaguchi}
\email{kawaguchi@eagle.sophia.ac.jp}
\affiliation{%
Department of Physics, Sophia University, Tokyo 102-8554, Japan
}
\author{Yasuki Tachibana}
\email{yasuki.tachibana@mail.ccnu.edu.cn}
\affiliation{
Institute of Particle Physics and Key Laboratory of Quark and Lepton Physics (MOE),
Central China Normal University, Wuhan, 430079, China 
}
\affiliation{%
Department of Physics, Sophia University, Tokyo 102-8554, Japan
}
\author{Tetsufumi Hirano}
\email{hirano@sophia.ac.jp}
\affiliation{%
Department of Physics, Sophia University, Tokyo 102-8554, Japan
}

\date{\today}

\begin{abstract}
We propose a new approach to initialize the hydrodynamic fields 
such as energy density distributions and four flow velocity fields 
in hydrodynamic modeling of high-energy nuclear collisions 
at the collider energies.
Instead of matching the energy-momentum tensor or putting 
the initial conditions of quark-gluon fluids at a fixed initial time, 
we utilize a framework of relativistic hydrodynamic equations with
source terms to describe the initial stage. 
Putting the energy and momentum loss rate of the initial partons 
into the source terms, we obtain hydrodynamic initial conditions 
dynamically. 
The resultant initial profile of the quark-gluon fluid 
looks highly bumpy as seen 
in the conventional event-by-event 
initial conditions. 
In addition, initial random flow velocity fields  also
are generated as a consequence of momentum deposition from the initial partons. 
We regard the partons that survive 
after the dynamical initialization process as the mini-jets 
and find sizable effects of both mini-jet propagation in the quark-gluon fluids 
and initial random transverse flow
on the final momentum spectra
and anisotropic flow observables.
We perform event-by-event $(3+1)$-dimensional ideal hydrodynamic simulations with 
this new framework that 
enables us to describe 
the hydrodynamic bulk collectivity, 
parton energy loss, and interplay among them in a unified manner.

\end{abstract}


\maketitle

\section{Introduction}
High-energy nuclear collision experiments
are performed at the Large Hadron Collider (LHC) at CERN
and at the Relativistic Heavy Ion Collider (RHIC)
at Brookhaven National Laboratory (BNL)
towards the understanding of bulk and transport
properties of the deconfined nuclear matter, the quark gluon plasma (QGP) \cite{Yagi:2005yb}. 
One of the major discoveries at the RHIC 
is large azimuthal anisotropy
\cite{Ackermann:2000tr,Adler:2001nb,Adler:2002pb,Adler:2002pu,Adams:2004bi,Adcox:2002ms,Adler:2003kt,Back:2002gz,Back:2004zg,Back:2004mh}, 
which is comparable with the results 
 from relativistic hydrodynamic models 
\cite{Kolb:2000fha,Kolb:2001qz,Teaney:2000cw,Hirano:2001eu,Hirano:2002ds}. 
This triggers a lot of further theoretical and experimental efforts
to deeply understand the QGP fluids.

Hydrodynamic modelings have been greatly developed 
as the most standard approach to describe the soft dynamics 
in high-energy nuclear collisions 
\cite{ Bozek:2009dw,Song:2010mg, Schenke:2010rr, 
Gardim:2011xv,Bozek:2011ua, Qiu:2011iv, Qiu:2011hf, Gale:2012rq,
 DelZanna:2013eua,Niemi:2015qia,Pang:2016igs}. 
Especially, one of the most advanced models 
is to combine 
fully (3+1)-dimensional dissipative hydrodynamics 
with some initialization models and hadronic cascade codes so that
one describes the whole stage from event-by-event
initial hydrodynamic states to a final hadronic afterburner \cite{Ryu:2015vwa,Denicol:2015bnf,Karpenko:2015xea,Knospe:2015nva,Murase:2016rhl}.
This model enables one to extract transport properties of 
the QGP fluids from a variety of experimental data.
However 
how the system
reaches local thermal equilibrium rapidly after the first contact, 
which is suggested from the success of
hydrodynamic modeling, is not well understood yet. 
This is a longstanding open problem
and its solution is highly demanded.

Although the hybrid type model mentioned above has been successful
in the description of the bulk dynamics in high-energy nuclear collisions,
the model is applicable only in the soft sector, i.e., 
in the low transverse momentum ($p_{T}$) regions. 
On the other hand, hard partons
are produced together with the QGP at the collider energies
and are subject to lose their energy and momentum
as they traverse the bulk medium. 
To quantify the amount of energy loss in the dynamically 
evolving QGP, 
realistic solutions of relativistic 
hydrodynamic equations are utilized 
in Refs.~\cite{Hirano:2002sc, Hirano:2003hq,Hirano:2003pw}.
These studies aimed at describing 
hydrodynamic bulk collectivity 
in the low-$p_{T}$ regions 
and jet quenching in the high-$p_{T}$ regions
at once. At that time, the lost energies are so small that
these are not expected to change the bulk dynamics dramatically.
Later the contribution of the medium response to the jet propagation 
has been focused in the studies of jet substructure 
and recognized to be very important 
\cite{Tachibana:2014lja,He:2015pra,Casalderrey-Solana:2016jvj,Wang:2016fds}. 
In particular, very recently, 
it turns out that 
enhancement of yields observed 
at the large angles from the jet axis \cite{Khachatryan:2016tfj} is interpreted quantitatively 
from hydrodynamic responses to the jet propagation \cite{Tachibana:2017syd}. 
This suggests that 
the interplay between soft collective dynamics and hard jet propagation becomes more and more important. 

In addition to the jets with $p_{T} \gtrsim 100$ GeV,
a large number of mini-jets with $p_{T} \sim 2$-$10$ GeV also  are produced,
in particular, at the LHC energies.
With this line of thought, there is a very first attempt
to include propagation of \textit{multiple} mini-jets in the QGP fluids
and to study its consequences in flow observables \cite{Schulc:2014jma}.
In Ref.~\cite{Schulc:2014jma}, 
the optical Glauber model is employed for the initialization
of hydrodynamic fields,
and the resultant initial energy density distributions are smooth functions.
\footnote{The authors of Ref.~\cite{Schulc:2014jma}
 investigated the ``hot spot'' scenario
in which hard partons instantaneously deposit their energy to create a
hot spot on top of a smooth background. However, this 
is different from what we address in the following sections
since the profile of the background medium also fluctuates.}
This means that the higher order anisotropy $v_{n}$ ($n>3$)
 is induced mainly by the disturbance of the QGP fluids due to mini-jet propagation.
Although one is able to directly extract the effect of mini-jet propagation
on flow observables,
 there must be a complicated interplay among various mechanisms
which generate azimuthal anisotropy. These should be non-linear effects
that the final $v_{n}$ is not the sum of all possible effects.
One dominant mechanism of generating azimuthal anisotropy
should be fluctuating initial profiles from event to event.
Since the origin of mini-jets and that of hot spots of the medium
in fluctuating initial profiles
are the same, i.e.,  nucleon-nucleon inelastic collisions at the first contact
of two nuclei,
it is not obvious to divide soft (the medium) from hard (the mini-jets)
components.
This, in turn, demands a framework to treat soft and hard components
at once, both in the initial states and during the evolution of the system.
Thus, the main purpose of this paper
is to develop the first dynamical model to
do so within hydrodynamic modeling.
As an application of this model, we investigate the effects of
mini-jet propagation on
transverse-momentum spectra and anisotropic flow
in Pb+Pb collisions at the LHC energy.

The paper is organized as follows. In Sec.~II, we explain
a new approach to generate hydrodynamic fields.
We show results of transverse momentum spectra, elliptic flow,
and triangular flow coefficients in Pb+Pb
collisions at the LHC energy
in Sec.~III.
Section IV is devoted to a summary of the paper.

We use the natural unit, $\hbar = c = k_{B} =1$, and the Minkowski metric,
$g^{\mu \nu} = \mathrm{diag}(1, -1, -1, -1)$, 
throughout this paper.

\section{Model}
\label{sec:model}

We formulate a model in which all the matters produced in high-energy nuclear collisions 
are supposed to arise from the partons created at the first contact of two nuclei. 
We generate the partons 
by estimating their production from 
 an event generator \pythia\;\cite{Sjostrand:2007gs}
combined with
the Monte Carlo version of the Glauber model (MC-Glauber) \cite{Miller:2007ri}.
Then, these partons start to travel through the vacuum after their production 
and are assumed to lose their energy and momentum 
until either the preassigned hydrodynamic initial time or the time when 
their energies vanish completely. 
In the meantime, all the lost energy and momentum are put 
into the source term of hydrodynamic equations 
to generate medium fluids gradually. 
After the hydrodynamic initial time, 
the dynamics is the same as a conventional
QGP fluid + jet approach \cite{Tachibana:2014lja,Tachibana:2015qxa,Tachibana:2017syd}: 
Surviving partons are regarded as mini-jets
and deposit their energy and momentum
while traversing the fluids 
until their energies vanish completely or they escape from the fluids. 
The fluid evolves under the influence of the mini-jet propagation 
according to the hydrodynamic equations with source terms. 
Finally, the particle spectra from the fluids are calculated 
via the Cooper-Frye formula \cite{Cooper:1974mv}.

\subsection{Distribution of partons in phase space}
\label{sec:initial}

We first use the MC-Glauber model to estimate the number
of participants and that of binary collisions
at some transverse point $\bm{x}_\perp$.
For each pair of binary collisions, 
we run \pythia\;for one inelastic $p$+$p$ collision 
with switching off fragmentation.
Note that we neglect possible iso-spin effects in the initial collisions.

An incoherent sum of \pythia\;results for pairs of all binary collisions
would mean that multiplicity scales with $N_{\mathrm{coll}}$. 
However, $N_{\mathrm{coll}}$ scaling is
anticipated only in high-$p_{T}$ regions.
On the other hand, the dominant source
of multiplicity is soft, namely, low-$p_{T}$ particles. 
To account for approximate $N_{\mathrm{part}}$ scaling
of multiplicity, and to demonstrate the idea of a
``rapidity triangle or trapezoid'' \cite{Brodsky:1977de,Adil:2005qn,Hirano:2005xf,Hirano:2012kj},
one needs to perform a rejection sampling from the particle ensemble in \pythia\,
\footnote{Hydrodynamic initial conditions which contain the idea
of a ``rapidity triangle or trapezoid'' have been extensively discussed in
Refs.~\cite{Hirano:2005xf,Hirano:2012kj}. In these studies,
rapidity distributions
in $p$+$p$ collisions were parametrized simply by using
a smooth function.
Whereas, in the present paper,
we employ \pythia\;for particle production.
Multiplicity and the longitudinal profile fluctuate in this model, which
is important, especially, in small colliding systems~\cite{Kawaguchi:2017idv}.
%: 
%K.~Kawaguchi, K.~Murase, and T.~Hirano, to appear in EPJ Web of Conferences
%and in preparation.
}.

From the MC-Glauber model,
we first pick up
two nucleons (say, $A$ and $B$, respectively)
undergoing a binary collision.
Transverse positions of nucleon $A$ (positive beam rapidity) 
and nucleon $B$ (negative beam rapidity) are
assumed to be $\bm{x}_{\perp, A}$
and $\bm{x}_{\perp, B}$, respectively.
We next count the number
of binary collisions for each nucleon: $N_{A}$ 
($N_{B}$) means the number of nucleons which are collided by
nucleon $A$ ($B$).
Then we perform a rejection sampling for partons 
with transverse-momentum $p_{T}=\sqrt{p_{x}^{2}+p_{y}^{2}}$
and rapidity $Y=(1/2)\ln[(E+p_{z})/(E-p_{z})]$
from \pythia\;with a momentum-dependent
acceptance function,
\begin{eqnarray}
\label{eq:weight}
w(p_{T}, Y) & = & w(Y) \times \frac{1}{2}
\left[1-\tanh\left(\frac{p_{T}-p_{T0}}{\Delta p_{T}} \right)\right] \nonumber \\
&+& 1\times \frac{1}{2}\left[1+\tanh\left(\frac{p_{T}-p_{T0}}{\Delta p_{T}} \right)\right], \\
w(Y) & = & 
\frac{Y_{b}+Y}{2Y_{b}}\frac{1}{N_{A}} 
+\frac{Y_{b}-Y}{2Y_{b}}\frac{1}{N_{B}}.
\end{eqnarray}
Here $Y_{b}$ is the beam rapidity, $p_{T0}$ is a parameter to divide soft and hard transverse-momentum regions, and $\Delta p_{T}$ is a width parameter for the crossover region.
We repeat this procedure for each pair of nucleons 
undergoing a binary collision in one heavy-ion collision event.
Notice that  the case of $N_{A} = N_{B} = 1$ 
results in an ordinary single \pythia\;event.

Let us suppose 
$N_{A}$ nucleons (with \textit{negative} beam rapidity) and $N_{B}$ nucleons
(with \textit{positive} beam rapidity)
are collided with each other.
It should be noted that $N_{A}$ means the number of nucleons collided by
nucleon $A$ which is going with \textit{positive} beam rapidity. 
In this case, the number of binary collisions is given by 
 $N_{\mathrm{coll}} = N_{A}N_{B}$
and we call $N_{A}N_{B}$ times inelastic $p$+$p$ events from \pythia.
Then for each two-dimensional bin of $p_T$ and $Y$ in each binary collision, 
we take samples of the partons by using the acceptance function~(\ref{eq:weight}). 
In the high-$p_{T}$ region,
the parton yields scale with 
\begin{equation}
N_{A} N_{B} w(p_{T}\gg p_{T0}, Y) \approx N_{A}N_{B}
\equiv N_{\mathrm{coll}}.
\end{equation}
Therefore, all the partons created in $N_A N_B$ times \pythia\, simulations remain 
at the high-$p_T$ limit in the entire rapidity region 
even after the rejection. 
Whereas,
in the low-$p_{T}$ region,
the parton multiplicity scales with 
\begin{eqnarray}
N_A N_B w(p_{T}\ll p_{T0}, Y) &\approx& N_A N_B w\left(Y\right) \nonumber\\
&=& 
\frac{Y_{b}+Y}{2Y_{b}}N_{B} +\frac{Y_{b}-Y}{2Y_{b}}N_{A}.
\end{eqnarray}
Thus, at the low-$p_T$ limit, 
the number of partons remaining after the rejection 
exhibits rapidity distributions
under the idea of a rapidity triangle or trapezoid: 
The parton yields at beam rapidity $Y= Y_{b}(-Y_{b})$ 
after the rejection 
reduce to 
those obtained from $N_{B}$ ($N_{A}$) times \pythia\, simulations.
At midrapidity $Y=0$, the multiplicity scales with
\begin{equation}
N_{A}N_{B}w(p_{T}\ll p_{T0}, Y=0) \approx \frac{N_{A}+N_{B}}{2} \equiv \frac{N_{\mathrm{part}}}{2}
\end{equation}
 as anticipated. 

\begin{figure}[htbp]
\vspace{12pt}
\begin{center}
\includegraphics[bb=0 0 360 254, width=0.45\textwidth]{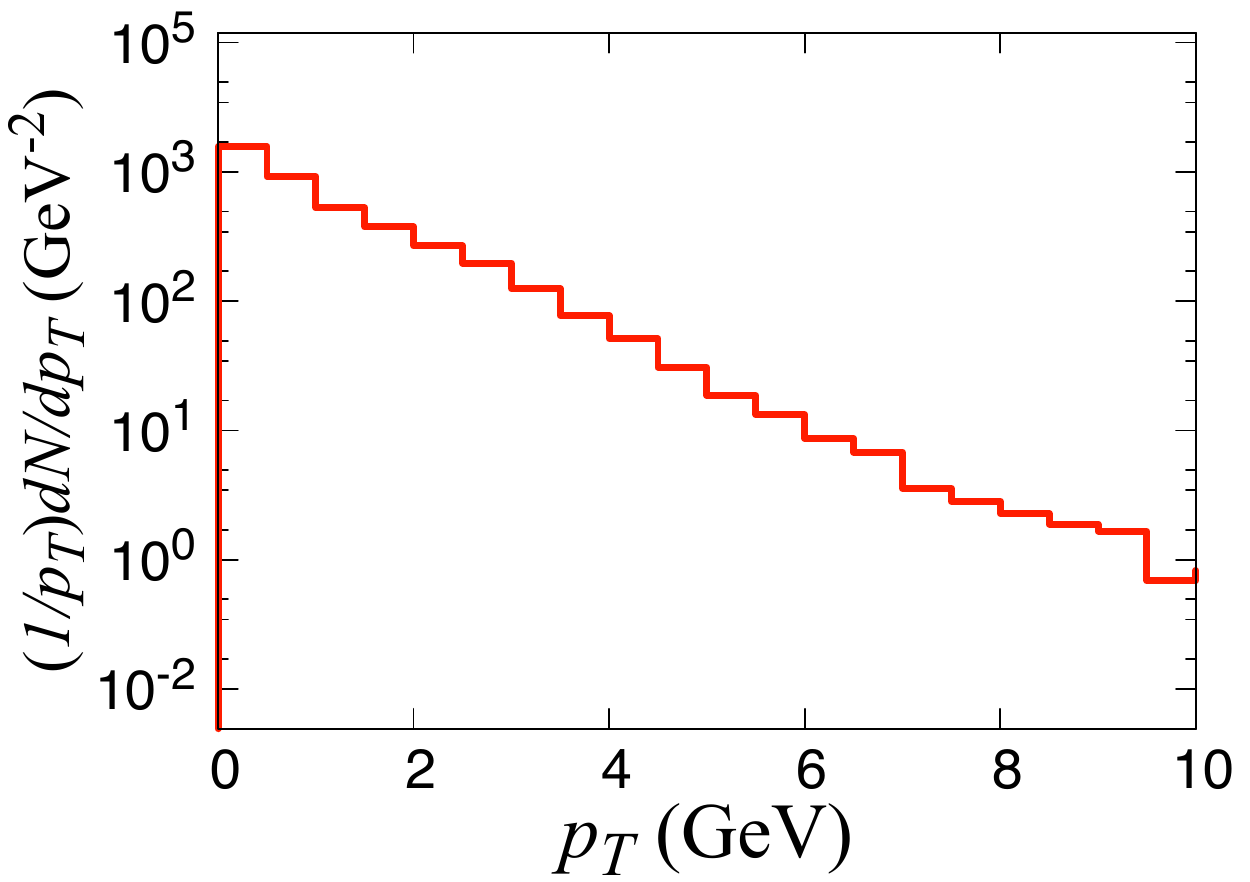}
\end{center}
\caption{(Color online) The $p_{T}$ spectrum of the  initial partons in one central ($b=0$ fm) Pb+Pb
collision at $\sqrt{s_{NN}}=2.76$ TeV.
The number of binary collisions is estimated to be $N_{\mathrm{coll}}=1983$
in this particular event. }
\label{fig:parton_pt_dist}
\end{figure} 
Figure \ref{fig:parton_pt_dist} shows 
a $p_{T}$ distribution of 
initial partons in one central ($b=0$ fm) 
Pb+Pb event at $\sqrt{s_{NN}}=2.76$ TeV. 
Here we choose parameters $p_{T0}=2$ GeV and $\Delta p_{T}=1$ GeV.
These parameters should have been determined through
comparison of the model results with, e.g., experimental data of 
nuclear modification factors $R_{AA}$.
However, such detailed but time-consuming analyses will be postponed for future work.
Hence we fix these parameters so as to obtain a smooth $p_{T}$ distribution in this paper.

Using transverse positions of the pair
of binary collision
and rapidity of a produced parton, 
positions of the produced parton $i$ in the configuration space
are determined to be
\begin{eqnarray}
\label{eq:hadron_transverse_position}
\bm{x}_{\perp}^{i}
& = & (x^{i}, y^{i})  =  \frac{\bm{x}_{\perp, A}+\bm{x}_{\perp, B}}{2}
+\frac{\bm{x}_{\perp, A}-\bm{x}_{\perp, B}}{2Y_{b}}Y^{i},\\
\label{eq:hadron_longitudinal_position}
\eta_{s}^{i} & = &Y^{i}.
\end{eqnarray}
Thus the positions in the configuration space
are	 determined mainly from the MC-Glauber model.
Transverse positions ($\ref{eq:hadron_transverse_position}$)
 slightly are shifted randomly
within  
the geometrical 
cross section in inelastic $p$+$p$ collisions, $\sigma_{\mathrm{in}}^{pp}$.
Here we implicitly assume that particle production occurs
in the transverse area of a hadron string or a color flux tube
which is elongated along Eq.~(\ref{eq:hadron_transverse_position})
in rapidity space.

In this way, we obtain phase-space distributions
of partons just after
collisions of high-energy nuclei 
$f(t=0^{+}, \bm{x}; \bm{p})$ 
on an event-by-event basis.

\subsection{Hydrodynamic equations with source terms}
\label{sec:dynamics}

Relativistic hydrodynamic equations with source terms
\begin{equation}
\label{eq:qgp+jet}
\partial_\mu T^{\mu \nu} = J^\nu
\end{equation}
have been used extensively in the physics of jet quenching and its
effects on medium \cite{Tachibana:2014lja,Tachibana:2015qxa,Tachibana:2017syd}.
Here $T^{\mu \nu}$ is the energy-momentum tensor of the fluids, and
$J^\nu$ is the source term for the fluids.
In this paper, we neglect possible dissipative effects, which brings
the energy-momentum tensor to be 
$T^{\mu \nu}=(e+P)u^{\mu \nu}-Pg^{\mu \nu}$, 
where $e$ is the energy density, $P$ is the pressure, and $u^{\mu}$ 
is the four-fluid velocity.
We employ an equation of state from a recent
lattice QCD result \cite{Borsanyi:2013cga}.
Here it is assumed that energy and momentum deposited from partons
are instantaneously equilibrated. 
We utilize this framework to generate initial hydrodynamic fields
dynamically.

We introduce two time scales, $\tau_{00}$ and $\tau_{0}$.
At $\tau=\sqrt{t^2-z^2}= 0$ fm, the highly Lorentz-contracted two nuclei are collided with each other at $z=0$ fm
and partons are produced as discussed in the previous subsection.
Until $\tau_{00}$, all partons are supposed to be formed.
Thus $\tau_{00}$ can be regarded as the formation time of the partons.
From $\tau_{00}$ to $\tau_{0}$,
all partons travel while losing their energy and momentum 
regardless of whether the medium exists at the position of the partons.
To do so,
we solve Eq.~(\ref{eq:qgp+jet}) with vanishing initial conditions of hydrodynamic
fields $T^{\mu \nu}(\tau=\tau_{00}) = 0$ until $\tau_{0}$ 
by modeling the source terms: 
\begin{eqnarray}
J^\mu(x) & = &\sum_{i}J_{i}^{\nu}(x), \\
J_{i}^\nu (x)& = & -\frac{dp_{i}^{\mu}}{dt}\delta^{(3)}(\bm{x}-\bm{x}_{i}(p_i,t)),\\
\bm{x}_{i}(t) & = & \bm{x}_{i}(t=0)+\frac{\bm{p}_{i}}{p_{i}^{0}}t.
\end{eqnarray}
Here the summation is taken over all partons.
In this paper, we assume a constant energy and momentum loss rate of  
$dE_i/dt = d|\bm{p}_i|/dt = 5$ GeV/fm
at $\tau_{00}<\tau<\tau_0$.
We solve hydrodynamic equations with source terms (\ref{eq:qgp+jet})
until $\tau=\tau_0$ to obtain hydrodynamic initial states in the
ordinary sense.

In the analysis of the observables, we have an option that all four fluid velocities
are reset to be the Bjorken scaling solution \cite{Bjorken:1982qr}
at $\tau=\tau_0$. 
This option mimics a conventional event-by-event Glauber-type initial condition 
which has a bumpy profile of matter density and no transverse flow
on the transverse plane.

\subsection{Freeze-out and flow coefficients}

After $\tau=\tau_0$,
we continue to solve Eq.~(\ref{eq:qgp+jet}) until the maximum temperature
goes below a fixed decoupling temperature of $T=T_{\mathrm{dec}}$.
The energy loss of the mini-jets in the QGP fluids after $\tau = \tau_0$ will be discussed 
in the next subsection.
To obtain the momentum distributions of hadrons from the medium 
after the hydrodynamic evolution, 
we use the Cooper-Frye formula \cite{Cooper:1974mv},
\begin{eqnarray}
\!\!\!E\frac{dN_i}{d^3p}
&=&
\frac{g_i}{\left(2\pi\right)^3}\!\!\int_{\!\Sigma}\!
\frac
{p^\mu d\sigma_{\mu}\left(x\right)}
{\exp\!\left[{p^{\mu}u_{\mu}\left(x\right)}/{T\left(x\right)}\right]\!\mp_{\rm BF}\!1}, \label{eqn:C-F}
\end{eqnarray}
where 
$g_i$ 
is the degeneracy, 
$\mp_{\rm BF}$ corresponds to 
Bose or Fermi statistics for hadron species $i$, 
and 
$\Sigma$ is 
the freeze-out hypersurface. 
Here the freeze-out is assumed to occur at a fixed temperature of $T_{\rm dec}=160\,{\rm MeV}$. 

Flow coefficients of azimuthal angle distributions 
at midrapidity $Y=0$,
are calculated from the event plane method. 
The event plane angle of the $n$-th order for the $j$-th event is obtained as 
\begin{eqnarray}
\tan n \Psi^j_n 
&=&
\frac
{\langle \sin n \phi_p \rangle_j}
{\langle \cos n \phi_p \rangle_j}, 
\end{eqnarray}
where $\phi_p$ is the azimuthal angle in the momentum space and
the angle bracket means the average over the particle ensemble 
at midrapidity $Y=0$ 
in a single event,
\begin{eqnarray}
\langle \mathcal{O} \rangle_j
&=&
\left.
\frac
{\int dp_T d\phi_p \mathcal{O} \frac{dN^{j}}{dp_T d\phi_pdY}}
{\int dp_T d\phi_p \frac{dN^{j}}{dp_T d\phi_pdY}}\right|_{Y=0}.
\label{eq:av}
\end{eqnarray}
Averaging over all events, we obtain 
the flow coefficients as a function of $p_T$ at midrapidity $Y=0$,
\begin{eqnarray}
\!\!\!\!\!\!\!\!
v_n(p_T)
\!&=&\!
\frac1{N_{\rm ev}}\!
\sum_j^{N_{\rm ev}}\!
\left.\frac
{\int\!\!d\phi_p \cos n(\phi_p \!- \!\Psi_n^j )\! \frac{dN^{j}}{dp_T d\phi_pdY}}
{\int\!\! d\phi_p \frac{dN^{j}}{dp_T d\phi_pdY}}\right|_{Y=0}\!\!.
\label{eq:vn}
\end{eqnarray}

In this paper, 
we calculate 
spectra and flow coefficients 
of charged pions directly emitted from 
the decoupling hypersurface. 
For more quantitative analyses, 
the effects of hadronic rescatterings on and
the contribution of the decays of hadron resonances 
and that of fragmentation of mini-jets to final spectra and flow coefficients 
are necessary  to be taken into account. 
We would like to leave them 
for a future study.

\subsection{Energy loss of mini-jets}
From $\tau = \tau_{0}$, we treat the surviving partons 
as mini-jets propagating through the medium. 
The mini-jets deposit their energy and momentum into the QGP fluid. 
We use the energy-loss rate for the mini-jets of the form \cite{Betz:2010qh}: 
\begin{eqnarray}
\frac{dE_i}{dt} = - \left[\frac{T\left(t, \BV[x]_i\left(t\right)\right)}
{T_0}\right]^3\left.\frac{dE}{dt}\right|_0, 
\label{Eq:energyloss}
\end{eqnarray}
where $T_0$ and 
$\left.dE/dt\right|_0$ are
the reference temperature 
and the reference energy-loss rate, respectively. 
In this paper, 
$T_0=500\,{\rm MeV}$ and $\left.{dE}/{dt}\right|_0=5\,{\rm GeV/fm}$ 
are chosen to give 
the typical values of the nuclear modification factor for high-$p_T$ particles ($p_T\sim 10\,{\rm GeV}/c$) 
in central Pb+Pb collisions at the LHC \cite{CMS:2012aa, Abelev:2012hxa}. 
The mini-jets continue to deposit their energy into the QGP medium 
according to Eq.~(\ref{Eq:energyloss}) 
until the local temperature drops to $T_{\rm dec} = 160~{\rm MeV}$ 
or until their energy vanishes. 
We assume that the mini-jet partons are massless 
and lose their momenta together with energy as
\begin{eqnarray}
\frac{d{\bm{p}_{i}}}{dt}=\frac{{\bm{p}_{i}}}{|{\bm{p}_{i}}|}\frac{d{{E}_{i}}}{dt}. 
\label{Eq:momentumloss}
\end{eqnarray}

\subsection{Remarks}

One is able to treat soft and hard physics in a unified manner 
in this model in principle.
At the moment, the model for the particle production and four-momentum loss
are quite simple. 
Although our dynamical initialization process would capture some aspects of 
local thermalization, we admit  this would not be the actual thermalization process. 
Nevertheless, our approach could be a first step toward constructing
a unified framework in full phase space based on relativistic hydrodynamics.
In the near future,
we will estimate final hadrons from mini-jets that survived in the final state 
 via fragmentation.
Thus we obtain the resulting spectra in the entire momentum region 
in high-energy nuclear collisions, 
starting from the partons created at the very initial stage.
Since soft and hard particles are treated at once,
correlations between soft and hard physics are naturally encoded in the framework. 

In the conventional Glauber-model-based initialization, 
energy and momentum are not conserved when hydrodynamic fields are set. 
The thermodynamic entropy density distribution of the medium 
is estimated 
from the number of participants, binary collisions, or produced particles. 
In this process, 
one often introduces an adjusting parameter of the multiplicity, 
or replaces particles with Gaussian functions with some smearing parameters. 
Also, initial flow velocity is often chosen from the Bjorken scaling solution \cite{Bjorken:1982qr} 
and is assumed
to vanish in the transverse direction.
In this case, the energy-momentum tensor does not match before and after initialization.\footnote{
Given the fact that a complete pre-thermalization model does not exist, 
the system is still far away from the local 
equilibrium state at the matching. 
In this case, the dissipative correction 
to an ideal part of energy-momentum tensor, 
$\delta T^{\mu \nu} = T^{\mu \nu} -T_{\mathrm{ideal}}^{\mu \nu}$,
must be huge so that dissipative hydrodynamics cannot be applicable. 
}.
On the other hand,
energy and momentum are conserved in the present framework all the way
through dynamical initialization.
There is no concept of matching the energy-momentum tensor
between pre-equilibrium physics and hydrodynamics
at the hydrodynamic initial time $\tau_{0}$,
no room for adjusting overall normalization of multiplicity,
or no smearing parameters from particles to hydrodynamic fields
in this model.
At $\tau = \tau_{0}$, the initial flow appears as a consequence of
 momentum deposition from initial partons.
Hence one does not need to parametrize the initial flow. 
Note that  the total energy is not exactly the same as the collision energy
in the current setting 
since we just combine 
the MC-Glauber model with \pythia.
In particular, the concept of $N_{\mathrm{coll}}$
implies the same $N_{\mathrm{coll}}$ times 
$p$+$p$ collisions at $\sqrt{s_{NN}}$ happen, which obviously overestimates
the total energy of a collision.
This could be resolved in principle by employing more sophisticated event generators. 
At present, our aim is to construct a framework of a unified approach in the entire phase space based on the Glauber picture. Therefore, the results obtained in this paper
should give a baseline of the follow-up studies.

\section{Results}
\label{sec:results}

\subsection{Initial states}
 
\begin{figure*}[htbp]
\begin{center}
\includegraphics[bb=0 0 374 288, width=0.45\textwidth]{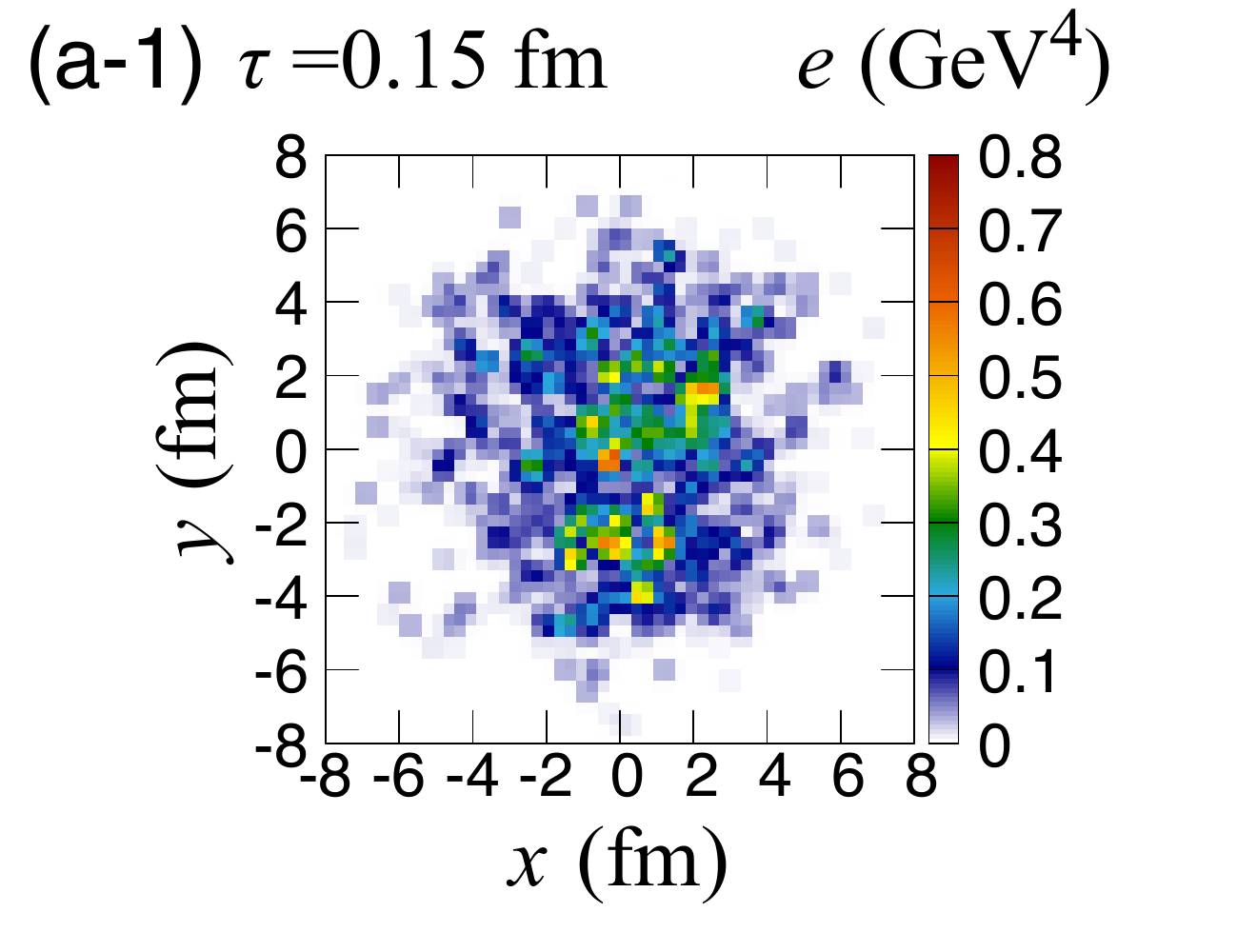}
\hspace{0.05\textwidth}
\includegraphics[bb=0 0 374 288,width=0.45\textwidth]{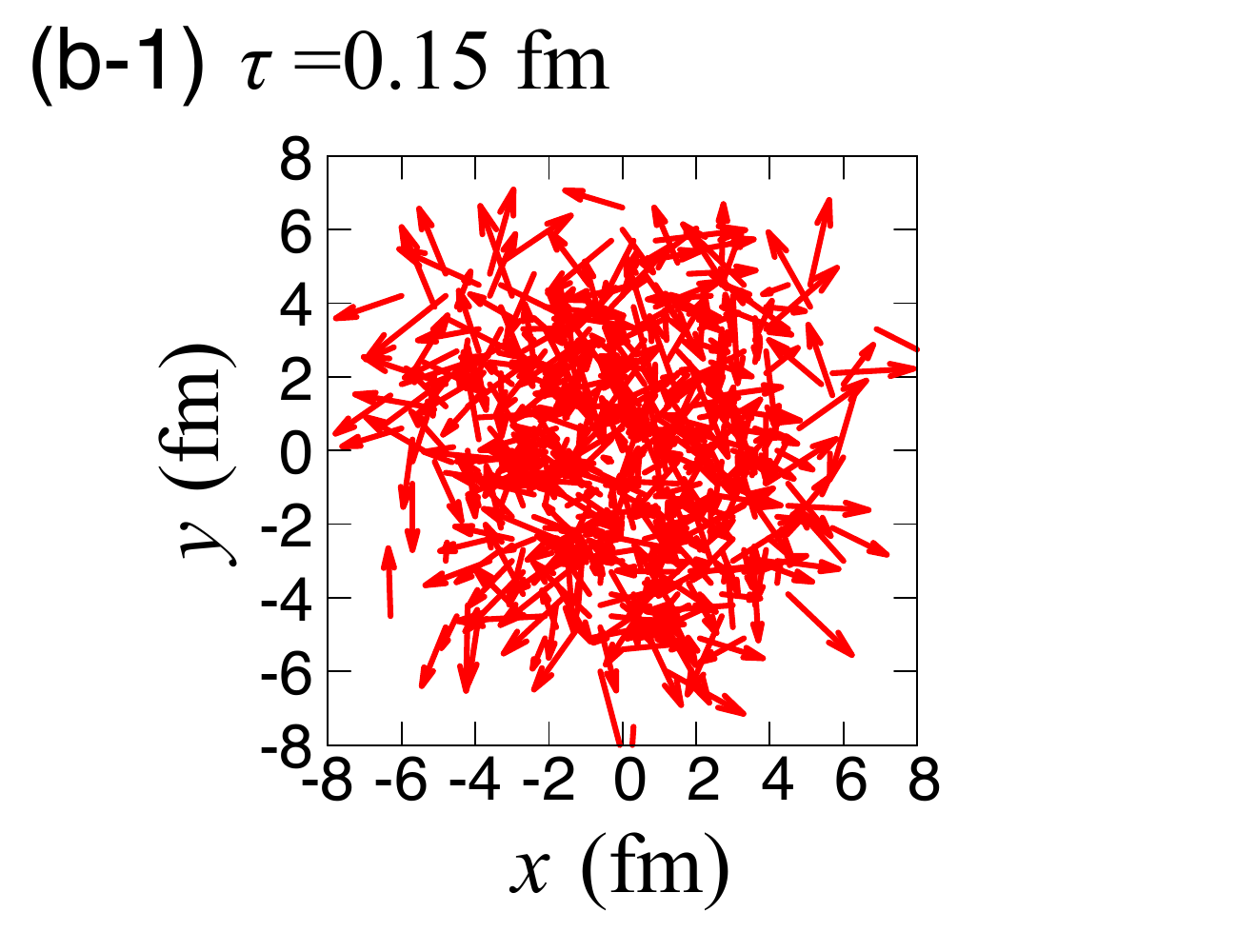}

\includegraphics[bb=0 0 374 288, width=0.45\textwidth]{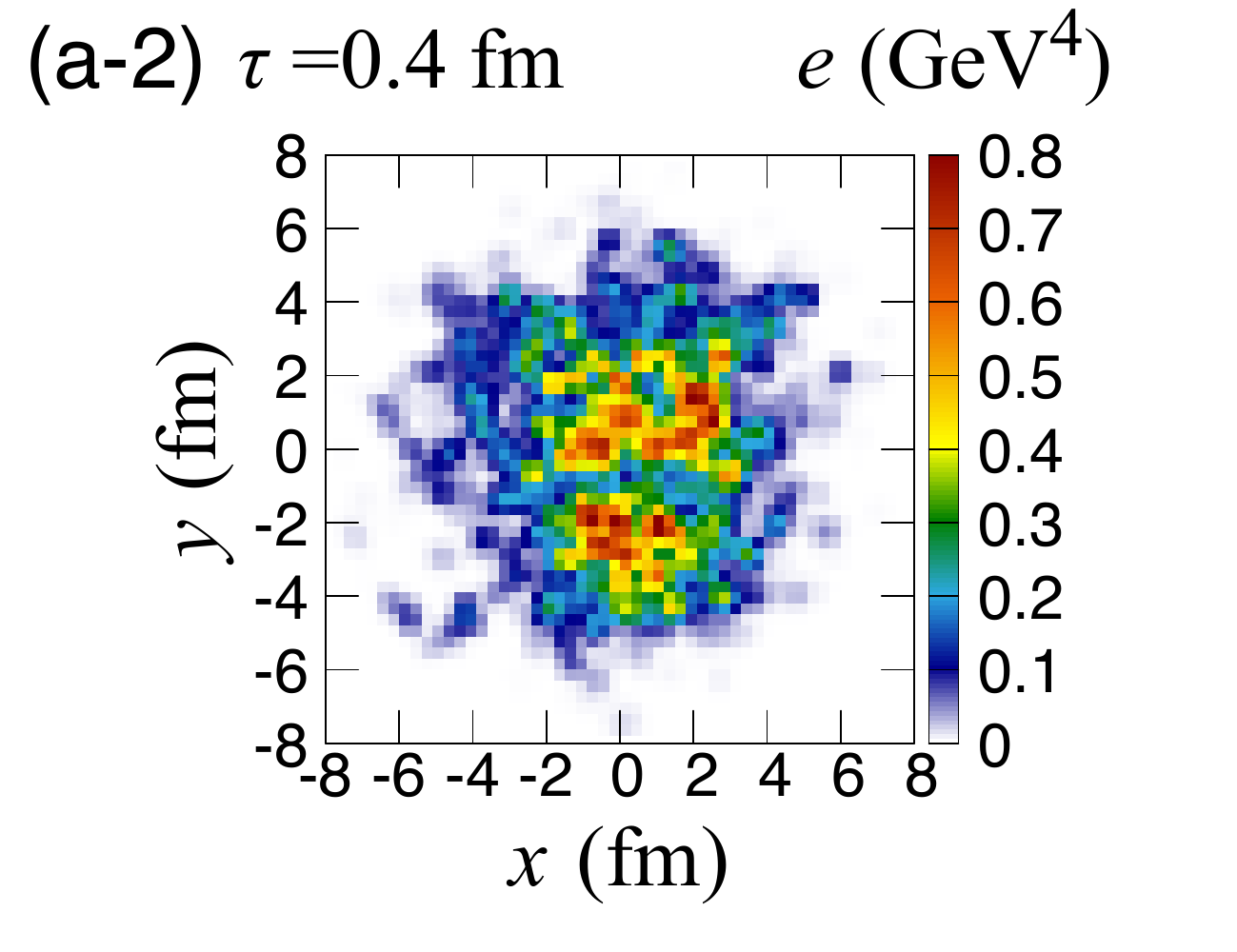}
\hspace{0.05\textwidth}
\includegraphics[bb=0 0 374 288,width=0.45\textwidth]{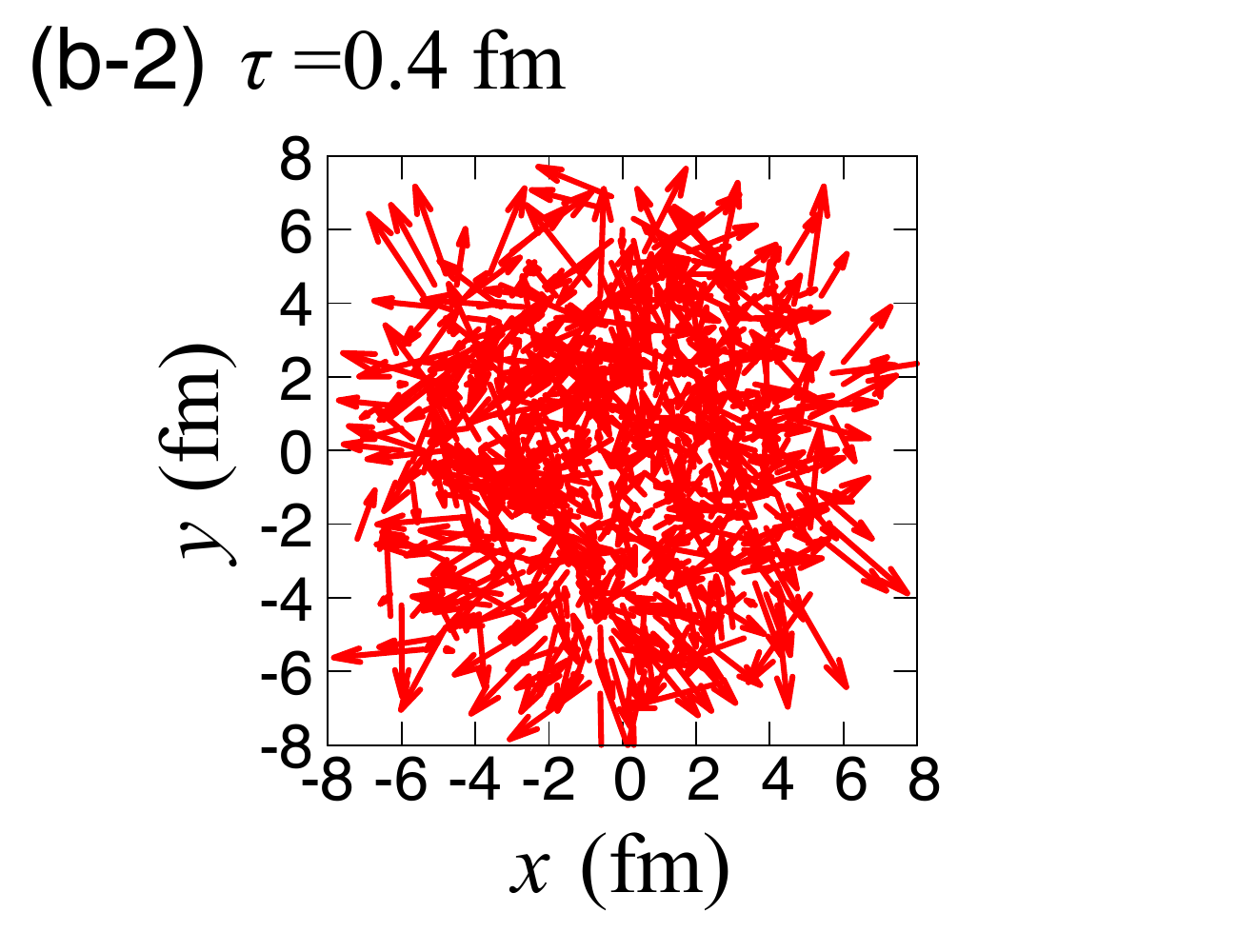}

\includegraphics[bb=0 0 374 288, width=0.45\textwidth]{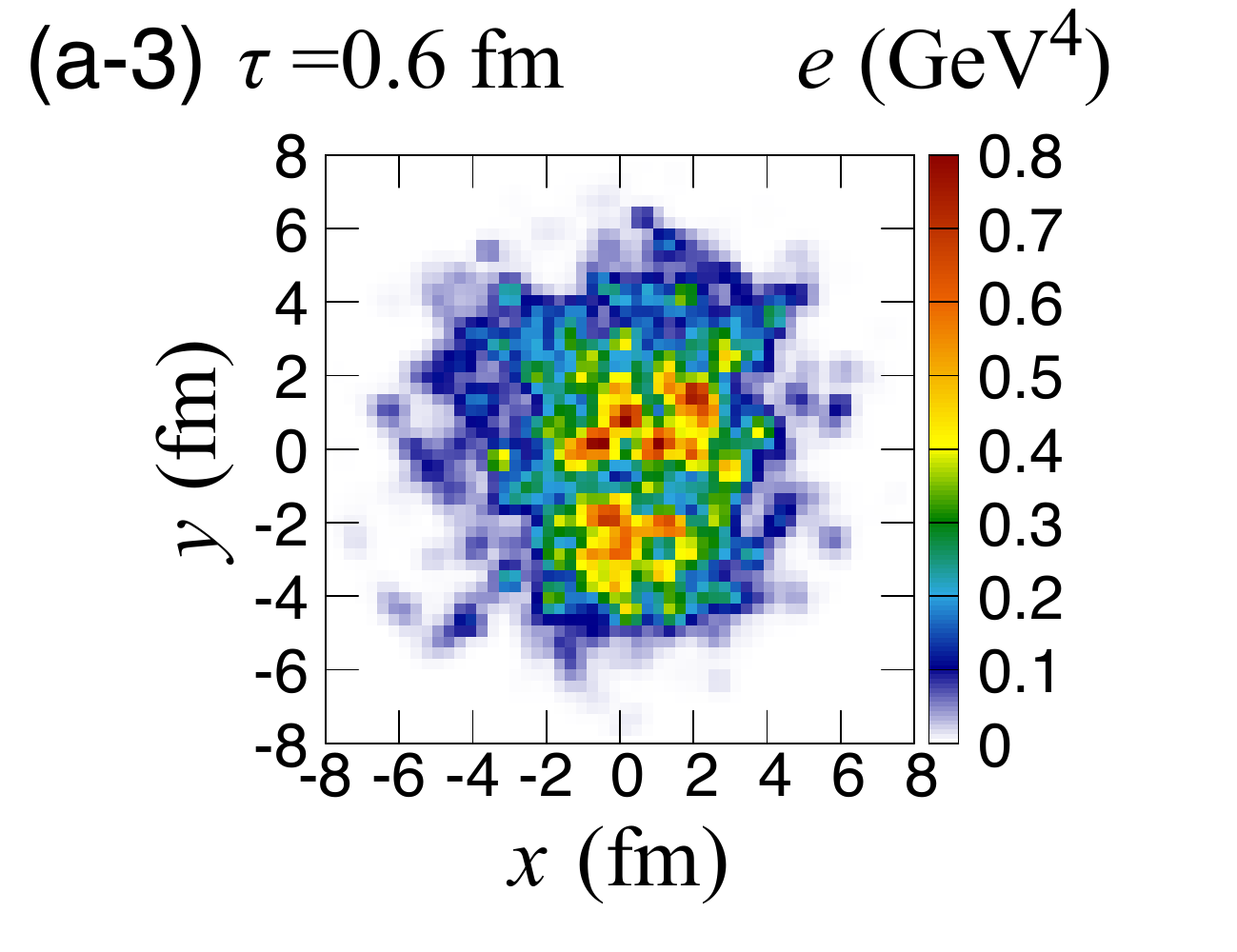}
\hspace{0.05\textwidth}
\includegraphics[bb=0 0 374 288,width=0.45\textwidth]{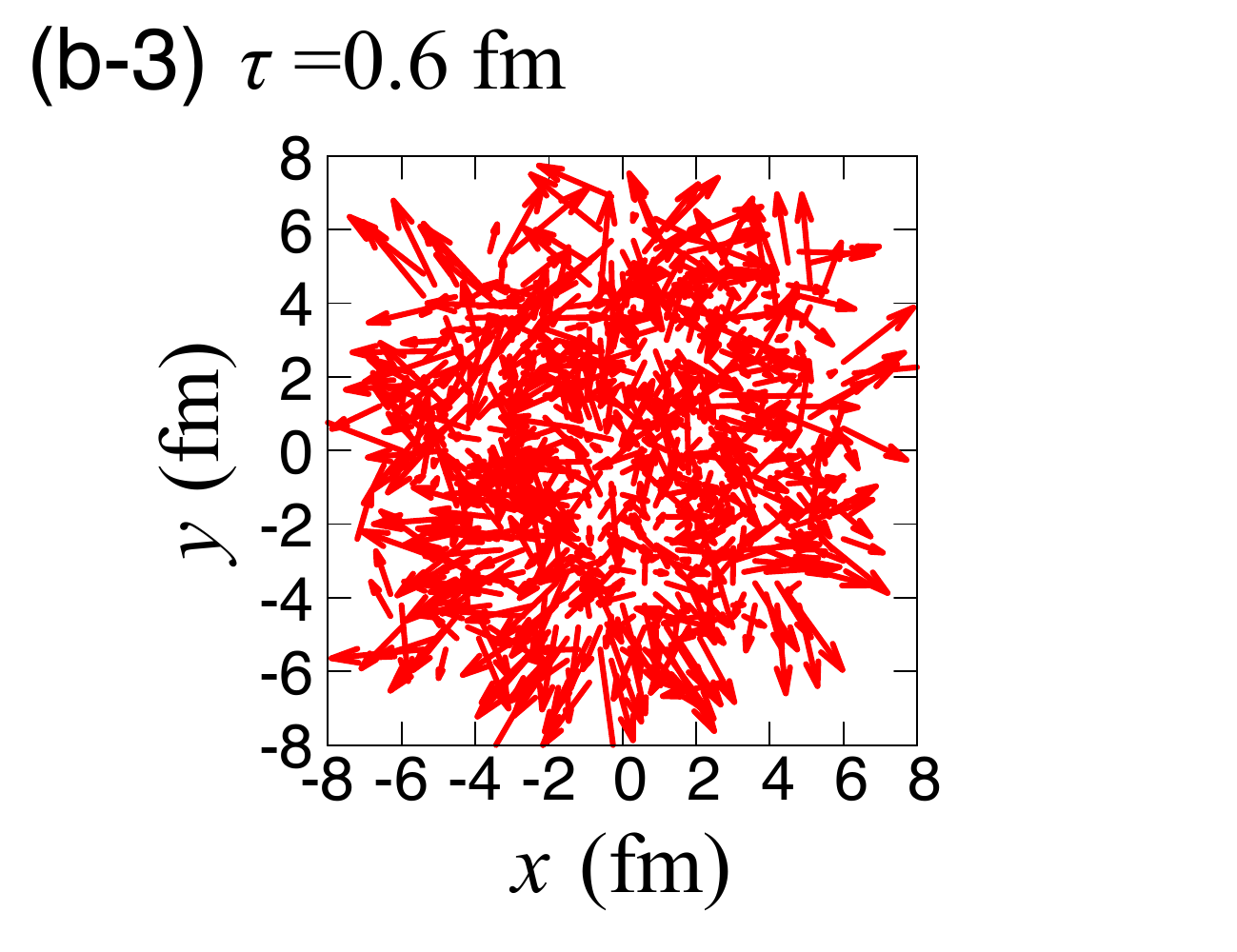}
\end{center}
\caption{(Color online)
Snapshots of the medium energy density distribution in units of GeV$^4$ (left panels)
and 
the transverse flow velocity (right panels), 
on the transverse plane 
in a Pb+Pb collision at the LHC energy.
From top to bottom, $\tau = 0.15$, $0.4$, and $0.6$ fm, respectively.
The impact parameter is $b=0.0$ fm for illustrative purposes.
}
\label{fig:snapshots}

\end{figure*} 
\begin{figure}[htbp]
\begin{center}
\includegraphics[bb=0 0 360 254,width=0.45\textwidth]{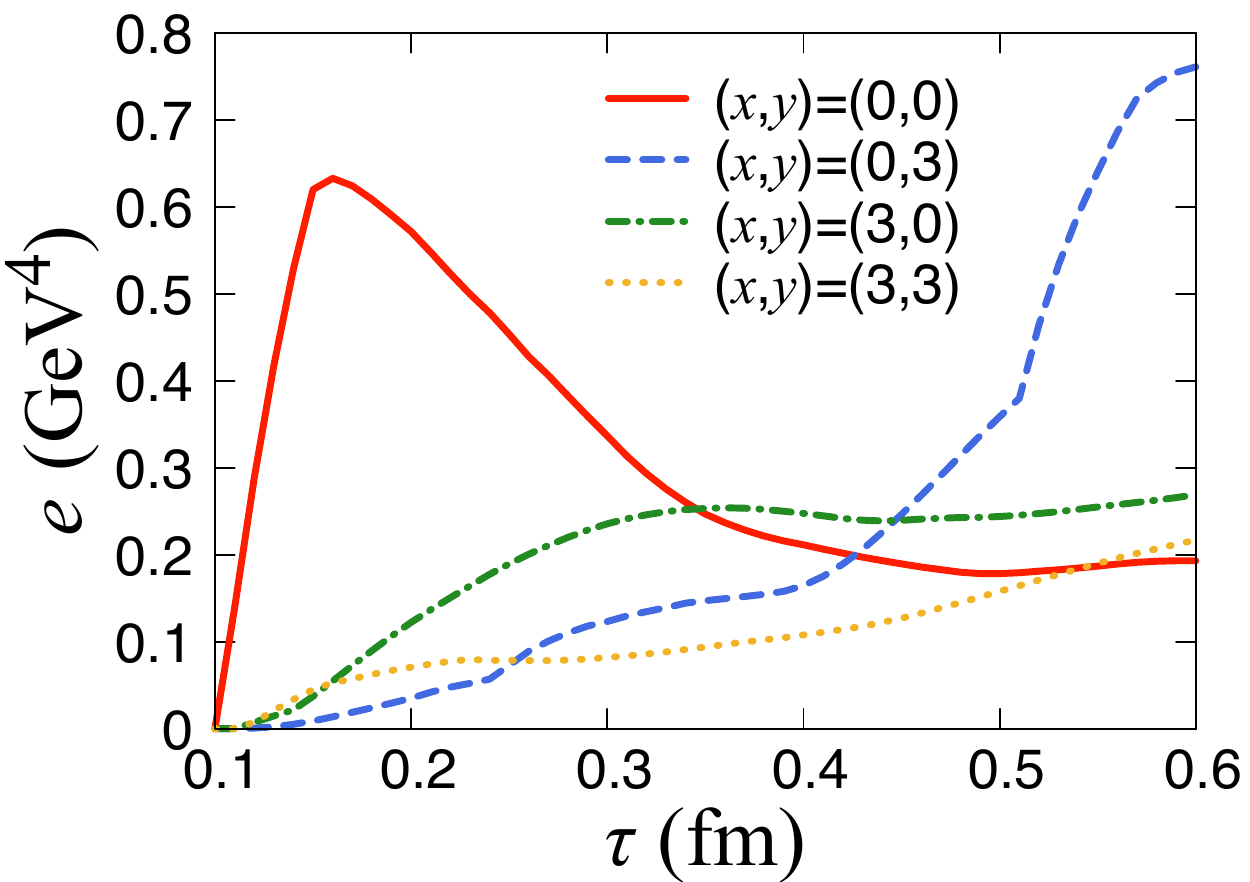}
\end{center}
\caption{(Color online)
An example of the time evolution of the energy density in units of GeV$^4$
at four representative transverse positions $(x, y)=(0, 0), (3, 0), (0, 3)$, and $(3,3)$ fm
during dynamical initialization $\tau_{00}< \tau < \tau_{0}$.  
}
\label{fig:e-tau}
\end{figure}

In the actual analysis, we choose $\tau_{00} = 0.1$ and
$\tau_{0} = 0.6$ fm throughout this paper.
Figure~\ref{fig:snapshots} shows 
snapshots of 
the medium energy density distributions (left panels)
and those of
the transverse flow velocity distributions (right panels) 
in a Pb+Pb collision at the LHC energy
during the dynamically generating process of 
the initial hydrodynamic fields. 
In these examples, 
the impact parameter is chosen to vanish for illustrative purposes.
The snapshots 
are taken on the transverse plane at $\eta_{s}=0$ at $\tau=0.15,\,0.4$,\, and
$0.6~{\rm fm}$. 
We see the gradual growth of the relatively higher-energy density area.
As partons lose their energies while traveling 
in vacuum or a medium, 
the medium energy density tries to increase rapidly.
On the other hand, the volume of the system expands in the longitudinal direction,
which would reduce the medium energy density on the transverse plane.
As a consequence, there is a competition between the growth 
and the dilution of the medium energy density during dynamical initialization.
For a fluid element at the fixed transverse position, some partons
come in during dynamical initialization
and deposit their energy. Thus energy density at that point suddenly increases.
The energy density can decrease as the ``hot spot'' expands radially.
Thus the time evolution of the energy density is a consequence of these various effects.
To see this more clearly, we show the time evolution of the energy density
at some fixed transverse positions in Fig.~\ref{fig:e-tau}.

The energy density profiles in the left panels of Fig.~\ref{fig:snapshots} 
look highly bumpy and 
quite similar to the conventional event-by-event initial conditions using the
MC-Glauber model.
However, a major difference between the present approach and the conventional
MC-Glauber type initialization is, as shown 
in the right panels of Fig.~\ref{fig:snapshots}, 
the appearance of random transverse flow
at $\tau=\tau_{0}$ 
which originates from random directions of the partons generated at the first contact.

\begin{figure}[htbp]
\vspace{12pt}
\begin{center}
\includegraphics[bb=0 0 360 254,width=0.45\textwidth]{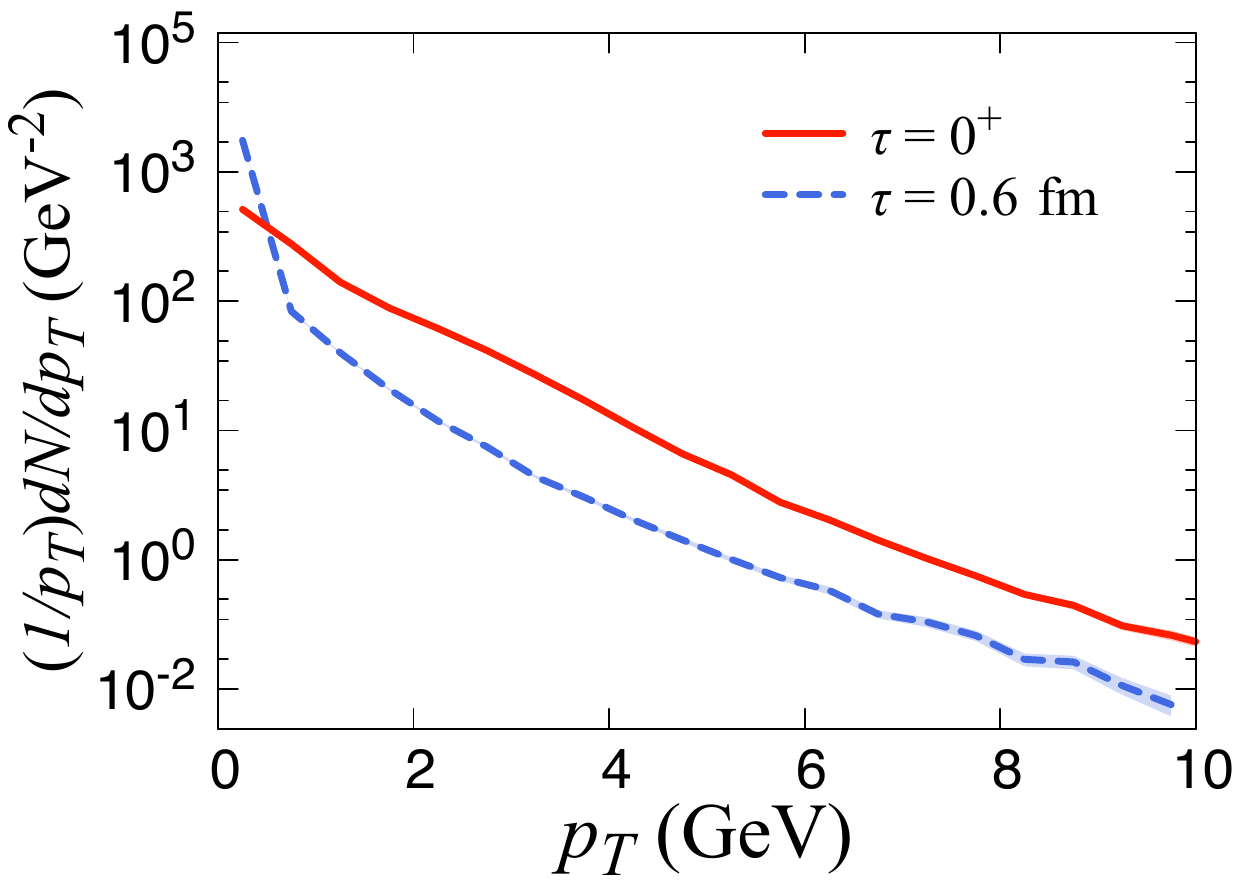}
\end{center}
\caption{(Color online)
The $p_{T}$ spectra of
the partons  at $\tau = 0$ (red solid line) and $\tau_0$ (blue dashed line).
 in a 30\%-40\% Pb+Pb collision at $\sqrt{s_{NN}}=2.76$ TeV.
The impact parameter is $b=9.32$ fm and the number of the event is $N_{\mathrm{ev}}=10^{2}$.}
\label{fig:dndpt_tau0}
\end{figure} 
Figure \ref{fig:dndpt_tau0} shows
the $p_{T}$ spectra of 
the partons  before and after dynamical initialization
 in 30-40\% Pb+Pb collisions at $\sqrt{s_{NN}}=2.76$ TeV. 
Hereafter the color bands in the figures represent the statistical errors. 
Here we fix an impact parameter of $b=9.32$ fm,
and the number of events is $N_{\mathrm{ev}}=10^{2}$.
It should be noted that 
a peak at the lowest-$p_{T}$ bin in the results at $\tau = \tau_0$ contains partons
with vanishing energy and momentum.
These surviving partons at $\tau = \tau_0$ traverse the medium as mini-jets.

\subsection{Spectra and flow}

\begin{figure}[htbp]
\vspace{12pt}
\begin{center}
\includegraphics[bb=0 0 360 254,width=0.45\textwidth]{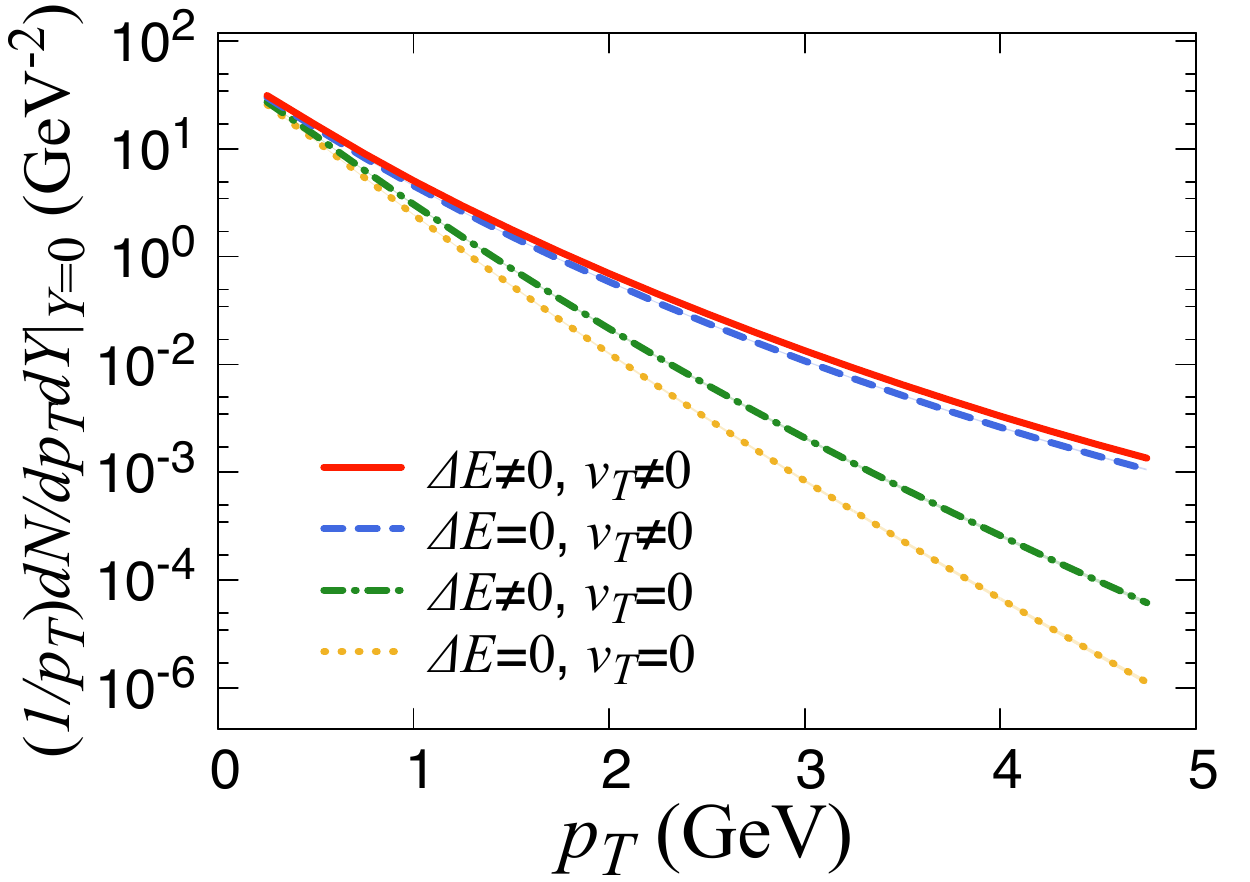}
\end{center}
\caption{(Color online)
The $p_{T}$ spectra of
charged pions  
 in a 40\%-50\% Pb+Pb collision at $\sqrt{s_{NN}}=2.76$ TeV 
with different settings for the initial flow and for the mini-jets' energy deposition. 
The impact parameter is $b=10.58$ fm and the number of the event is $N_{\mathrm{ev}}=10^{2}$.}
\label{fig:dndpt_final}
\end{figure} 
In Fig.~\ref{fig:dndpt_final},
the $p_{T}$ spectra of
charged pions  at midrapidity
in the four settings are compared with each other. 
Here we average over $N_{\mathrm{ev}}=10^{2}$ times 
Pb+Pb events at $b=10.58$ fm.
The $p_{T}$ spectrum looks like power-law behavior in our default setting
[$\Delta E \neq 0$, $v_{T}(\tau_{0}) \neq 0$] in which
random transverse flow appears at $\tau = \tau_{0}$ and, 
subsequently, the surviving partons
traverse the QGP medium as depositing their energy and momentum.
The $p_{T}$ spectrum is quite similar to the case when energy and momentum losses
are switched off after $\tau=\tau_{0}$ [$\Delta E = 0$, $v_{T}(\tau_{0}) \neq 0$].
The effect of mini-jet propagation is not significant in the $p_{T}$ spectra 
when the initial velocity is induced during 
the dynamical generation of the hydrodynamic fields. 
On the other hand, an exponential decrease with increasing $p_{T}$
is seen when transverse flow velocity is forced to vanish at $\tau=\tau_{0}$
[$\Delta E = 0$, $v_{T}(\tau_{0}) = 0$], 
which is consistent with the result from hydrodynamic models
with the conventional Glauber-type initial conditions.
As a consequence, the yields above $p_{T} \sim 3$ GeV in these cases
are much smaller than the ones with initial transverse flow.
This is the case even when 
mini-jets lose energy and momentum [$\Delta E \neq 0$, $v_{T}(\tau_{0}) = 0$].
This means the effect of initial random transverse flow is more 
significant than that of mini-jets propagation in the current setting. 

\begin{figure}[htbp]
\vspace{12pt}
\begin{center}
\includegraphics[bb=0 0 360 254,width=0.45\textwidth]{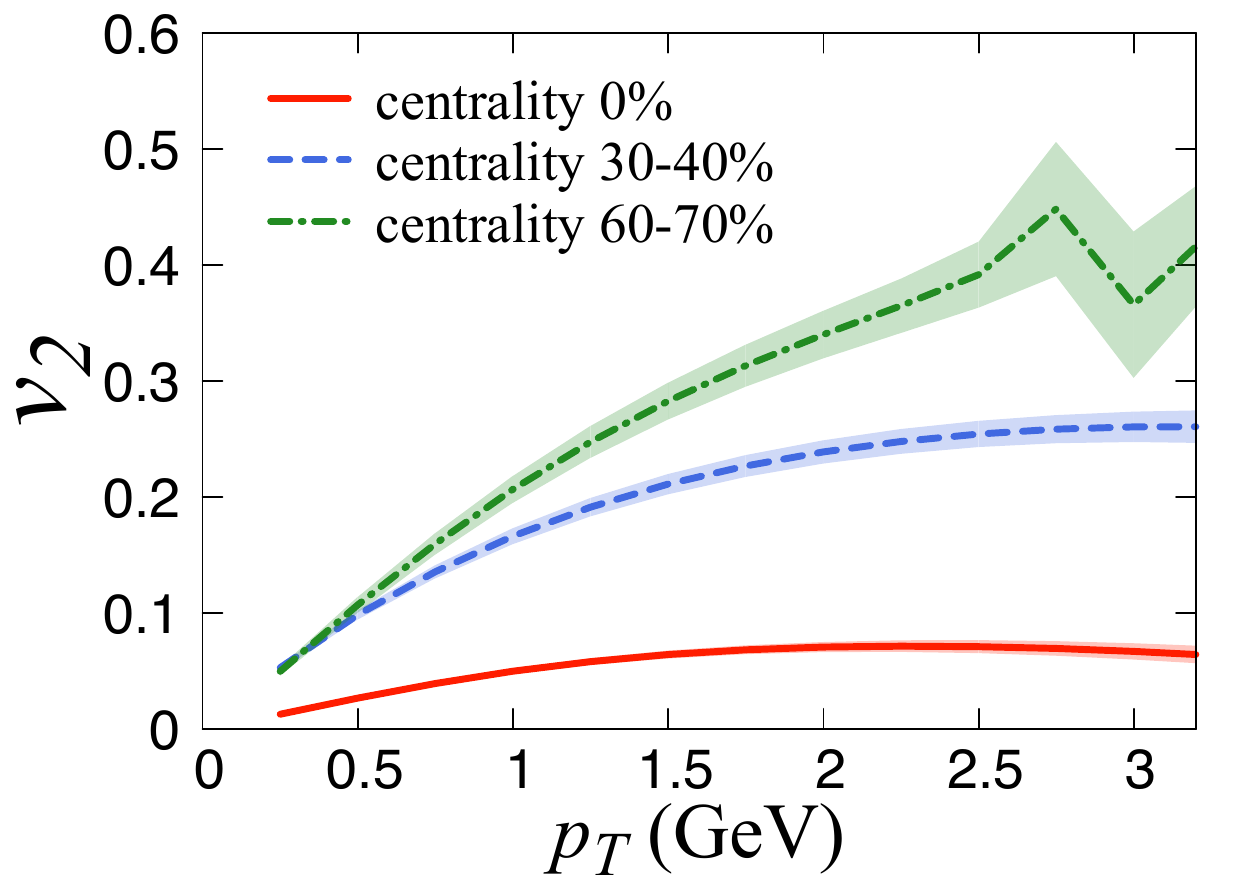}
\end{center}
\caption{(Color online)
The transverse-momentum dependence of elliptic flow parameter $v_2$ of
the charged pions 
at midrapidity $Y=0$ for centrality classes,
 0\%, 30\%-40\%, and 60\%-70\%,  
in Pb+Pb collisions at $\sqrt{s_{NN}}=2.76$ TeV.
}
\label{fig:v2_cent}
\end{figure} 
Shown in Fig.~\ref{fig:v2_cent} is the elliptic flow coefficient 
$v_2(p_T)$ of charged pions 
at midrapidity for different centrality classes 
with the default setting [$\Delta E \neq 0$, $v_{T}(\tau_{0}) \neq 0$] in our model. 
Here one can confirm 
the increase in $v_2$ with centrality, which is consistent with hydrodynamic results with conventional initialization. 
This means that 
the shape of the participants' region of nuclear collisions 
still has the largest contribution to $v_2$ 
even when there exist 
additional flow sources other than the initial pressure-gradient profile, i.e., 
the initial random transverse flow and the mini-jet induced flow. 
One also sees non-zero $v_{2}$ even at ``0\%'' 
centrality ($b = 0$ fm) 
caused by the event-by-event fluctuations of the initial profile and flow velocity of the medium. 

\begin{figure}[htbp]
\vspace{12pt}
\begin{center}
\includegraphics[bb=0 0 360 254,width=0.45\textwidth]{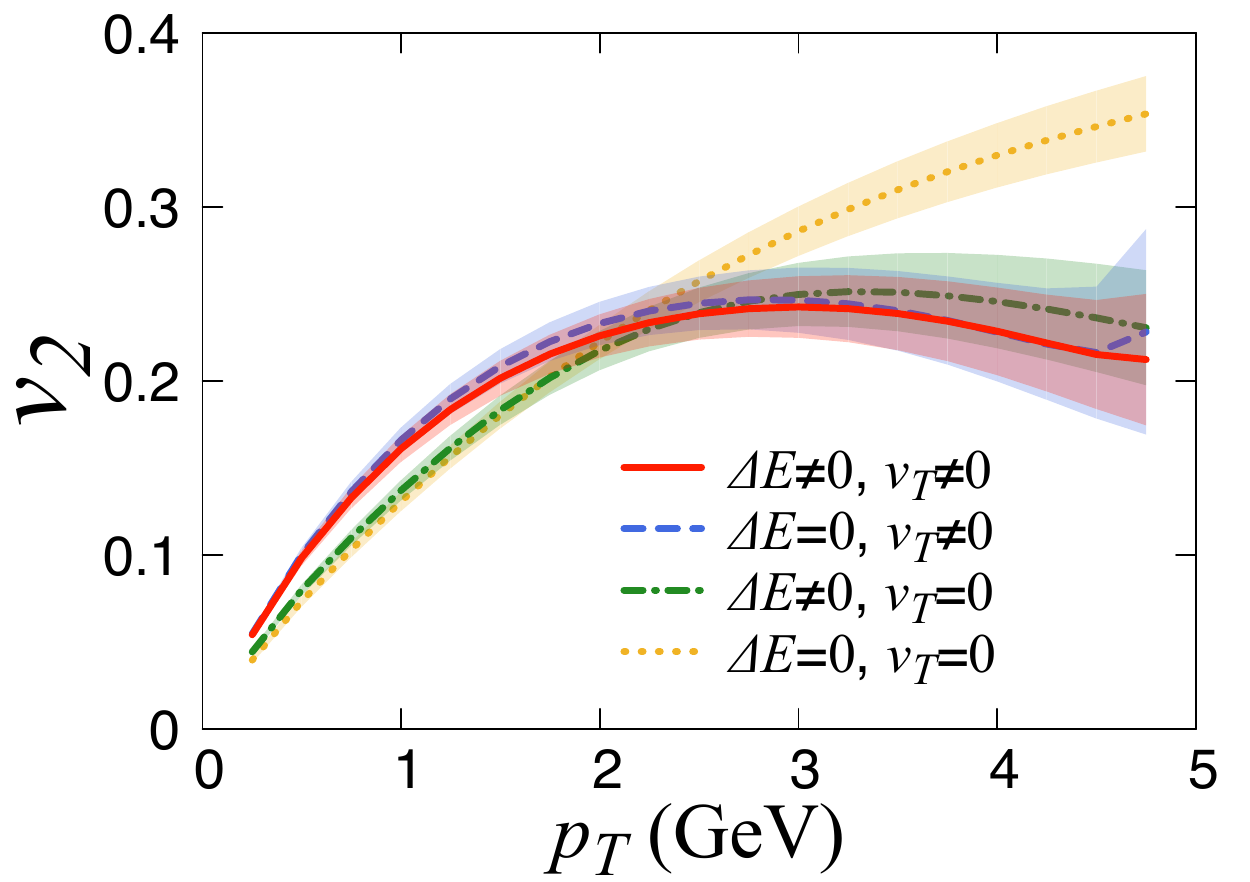}
\end{center}
\caption{(Color online)
The transverse-momentum dependence of elliptic flow parameter $v_2$ of
the charged pions at midrapidity $Y=0$ 
in 40\%-50\% Pb+Pb collision at $\sqrt{s_{NN}}=2.76$ TeV, 
with different settings for the initial flow and for the mini-jets' energy deposition. 
The impact parameter is $b=10.58$ fm, and the number of the event is $N_{\mathrm{ev}}=10^{2}$.
}
\label{fig:v2}
\end{figure} 
Figure~\ref{fig:v2} shows 
$v_2(p_T)$ of the charged pions 
at midrapidity 
in the four different settings in our model. 
In the low transverse momentum region of $p_T \sim 0$-$2$ GeV, 
one sees a little enhancement due to the initial flow fluctuations 
driven during the dynamical formation of the medium fluid. 
$v_2$ at high $p_T$ is dominated by the pions emitted from the medium 
with large flow velocity at freeze-out. 
When the initial flow velocity at $\tau_0$ and
the mini-jets energy loss are turned off [$\Delta E = 0$, $v_{T}(\tau_{0}) = 0$],
the medium flow velocity in the transverse direction is driven solely
by the initial pressure gradient in the medium at $\tau_0$. 
In particular, the region with large flow velocity in the medium strongly reflects
the initial profile of the medium and, as a result, $v_2(p_T)$ monotonically increases. 
On the other hand, when the initial random transverse flow velocity exists 
at $\tau = \tau_0$ and/or the mini-jets lose energy and momentum, 
the large flow velocity in the medium mainly is induced by the momentum deposition through the source term dynamically, and
it is not necessary to be aligned with that driven by the initial pressure gradient. 
As a result, the flow originating from the initial pressure gradient is disturbed 
and one can no longer see the monotonic increase in $v_{2}(p_{T})$
when there are additional flow sources other than the pressure gradient of the  initial profile. 
Regarding this point, it might be interesting to analyze 
factorization ratios $r_{2}(p_T^{a}, p_{T}^{b})$ \cite{Gardim:2012im}
for quantitative understanding of event plane decorrelation
in the transverse-momentum direction
within the current framework. 

\begin{figure}[htbp]
\vspace{12pt}
\begin{center}
\includegraphics[bb=0 0 360 254,width=0.45\textwidth]{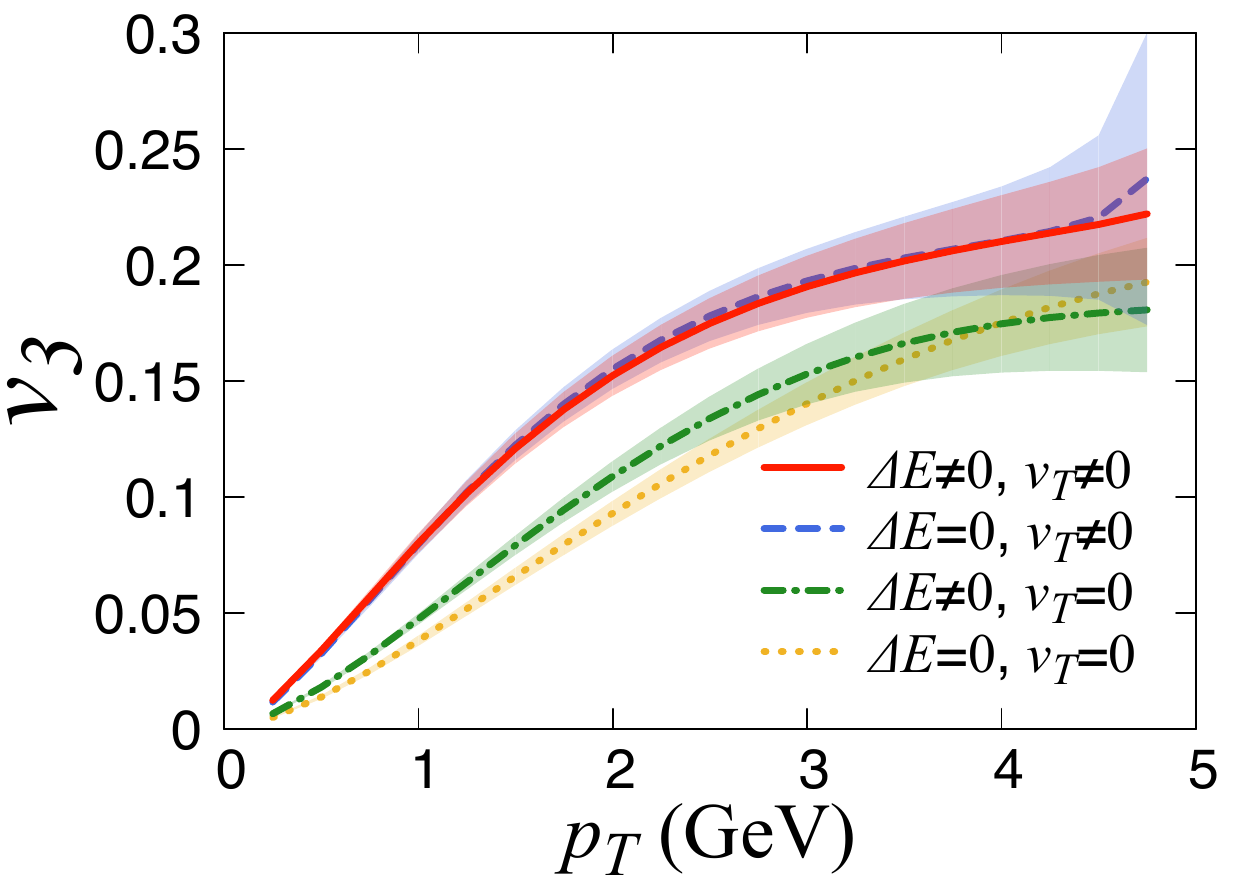}
\end{center}
\caption{(Color online)
The transverse-momentum dependence of triangular flow parameter $v_3$ of
the charged pions at midrapidity $Y=0$ 
 in 40\%-50\% Pb+Pb collision at $\sqrt{s_{NN}}=2.76$ TeV, 
with different settings for the initial flow and for the mini-jets' energy deposition. 
The impact parameter is $b=10.58$ fm, and the number of the event is $N_{\mathrm{ev}}=10^{2}$.
}
\label{fig:v3}
\end{figure} 
Figure~\ref{fig:v3} shows triangular flow coefficient 
$v_3(p_T)$ of the charged pions 
at midrapidity 
in the four different settings in our model. 
One sees that 
the initial flow fluctuations driven by the dynamical medium formation 
largely enhances the triangular flow. 
Triangular flow is induced mainly  by the initial fluctuations, 
and the global profile of the medium does not affect $v_{3}$ significantly. 
We also find the effect of mini-jet propagation on $v_{3}$ is not large.

\section{Summary}

In this paper, 
we formulated a new model 
to generate the medium evolving hydrodynamically 
in high-energy nuclear collisions. 
In the model, 
all the matters are supposed to be produced from partons created 
at the first contact of two nuclei. 
Combining \pythia\;with the MC-Glauber model, 
we generated these partons from 
incoherent $N_{\mathrm{coll}}$-times inelastic $p+p$ collisions. 
Then we applied the rejection sampling  
to obtain the initial phase-space distribution of the partons 
which satisfies 
$N_{\mathrm{coll}}$ scaling at high $p_T$, 
$N_{\mathrm{part}}$ scaling at low $p_T$, 
and exhibits rapidity triangle or trapezoid shape in the longitudinal direction
 at some transverse position. 
During the propagation through the vacuum after the production, 
the partons deposit their energy until the hydrodynamic initial time. 
The deposited energy is used to form the medium fluid dynamically 
via the source term of hydrodynamic equations. 
As well as the energy, the momentum also is deposited 
and, as a result, 
the medium fluid naturally acquires the initial flow velocity 
other than that driven purely from the initial pressure gradient. 
After the hydrodynamic initial time, 
we regarded the surviving partons as mini-jets traversing the medium fluid. 
The space-time evolution of the QGP with the energy and momentum deposited by the mini-jets 
also was  described by the hydrodynamic equations with source terms. 
To obtain the momentum distribution of the charged pions at freeze-out, 
we employed the Cooper-Frye formula. 

First, 
we saw how the medium and flow velocity 
are formed 
during the dynamical generation of the 
initial hydrodynamic field. 
The rapid energy-momentum deposition 
and the expansion of the system 
compete with each other, and, 
as a consequence, 
the energy density of the medium gradually grows. 
Then 
we investigated 
$p_T$ spectra and 
flow coefficients of azimuthal angle distributions 
by using the model. 
In particular we focused on the effects of 
the initial flow driven during the dynamical formation of the hydrodynamic field 
and 
flow induced by the mini-jet propagation. 
The initial random flow velocity makes the spectrum 
harder at high $p_T$. 
Although the momentum deposition also gives a similar contribution, 
it is not so significant. 

Next, we investigated 
the centrality dependence of the elliptic flow coefficient $v_2(p_T)$ 
in our model. 
$v_2$ increases with centrality as the hydrodynamic calculations with conventional initial-condition models, 
which indicates that the initial global shape of the medium also 
dominantly affects $v_2$ in our model. 
To study how 
the initial random flow velocity driven during the dynamical initialization 
and mini-jet induced flow 
affect anisotropic flow, 
we calculated 
$v_2(p_T)$ and $v_3(p_T)$ of the charged pions 
at midrapidity
with different settings 
for the initial condition. 
Both the initial random flow velocity 
and the 
mini-jet-induced flow 
disturb 
the flow driven by the initial pressure gradient 
and, as a result, suppress 
$v_2$ at high $p_T$.
On the other hand, 
the significant enhancement of $v_3$ 
was seen in the case with the the initial random flow velocity. 
Thus, we
found 
the initial random transverse flow and the mini-jet induced flow
indeed 
cause 
sizable anisotropic flow in the QGP fluid.
This strongly suggests the conventional hydrodynamic interpretation of
flow data based solely on initial eccentricity
should be revisited by taking account of the corrections from mini-jet
propagation and initial random velocity fields.

\section*{Acknowledgement}
The authors are very indebted to H.~Hamagaki 
for fruitful discussions in the very early stage of this study. 
YT is grateful to Y. Hirono for 
helpful discussions regarding numerical implementations. 
YT also acknowledges the kind hospitality of 
the nuclear theory group 
at Lawrence Berkeley National Laboratory where parts of this paper were completed. 

\appendix
%\bibliography{mybibfile}
\bibliography{ref}

\end{document}